# Design and technology of high-power couplers, with a special view on superconducting RF

W.-D. Möller
Deutsches Elektronen-Synchrotron, Hamburg, Germany

**Abstract**
The high-power RF coupler is the connecting part between the RF transmission line and the RF cavity and provides the electromagnetic power to the cavity and the particle beam. In addition to this RF function it also has to provide the vacuum barrier for the beam vacuum. High-power couplers are one of the most critical parts of the RF cavity system in an accelerator. A good RF and mechanical design as well as a high quality fabrication are essential for efficient and reliable operation of an accelerator.

## 1 Introduction

Modern accelerators require not only ever increasing performance, i.e., high accelerating gradients, but also high operating efficiency and reliability. The technical demands on the key components of an accelerating machine such as the RF accelerating structures and their related parts are very high. One of the most crucial components of the RF accelerating system is, in addition to the accelerating structure itself, the high-power coupler (PC). It is known that the performance and liability limitations of an RF accelerating system very often come from the PC's malfunctions.

Because the PC has to cope with multiple functions and has to fulfil tight requirements, the design can be very complex. Not only the demand for high power transfer, but, in the case of superconducting technology, also for low power losses in cryogenic environments, and cleanliness are important. The PC fabrication includes the use of many different materials, joining and coating technologies and treatments. During the design phase and fabrication planning, the assembly and handling have also to be taken into account because they can have an influence on the system performance.

Over many years the great interest in and importance of the design and technology of PCs has been reflected in the many papers and tutorials at CERN Accelerator Schools, Workshops on RF Superconductivity [1–3], and other conferences [4, 5]. An extensive choice of references about PC for superconducting RF applications can be found in Ref. [6].

## 2 General considerations

The RF high PC is the connecting part between the RF transmission line and the RF resonator (cavity). It has to fulfil the following RF functions:

It has to transfer the power to the cavity and to the beam at sometimes very high power levels in pulsed or continuous-wave (CW) operation. It has to match the impedance of the klystron and RF distribution system to the beam loaded cavity. It has to take into account that there is a strong mismatch in absence of beam between the unloaded cavity and the generator. This leads to full reflection in case of superconducting (SC) cavities. The wasted power has to be minimized and possibly the PC has to allow the change of the match for different beam loading. The dimensions should not allow non-TEM modes. The PC could also be used for HOM damping.

The non-RF functions are as follows:

The PC provides a vacuum barrier for the beam vacuum. It should not contaminate the accelerator vacuum. Therefore an easy cleaning and clean assembly must be possible. In case of SC cavities the PC has to be clean according to ISO 4 (‘dust free’). In a cryogenic environment at a SC accelerating system the PC in addition provides the bridge between room and cryogenic temperature. Low static and dynamic thermal losses to the low-temperature (2–4 K) resonator and mechanical flexibility for the temperature cycles and thermal expansions have to be taken into account. Not a function, but also important are low fabrication costs.

Malfunction of the PC could lead to a power limitation by arcing, multipacting, or window heating and this limits the RF accelerator performance. Over long operation times, possible damage to the coupler can make the accelerating system inoperable. The most serious fault is a vacuum leak of the ceramic. This leads to a bad contamination of the accelerator vacuum system and, in case of SC RF, to a contamination of the very delicate SC resonator surface. The recovery is very time consuming and expensive.

## 3 Rectangular wave guide vs. coaxial coupler

There are two main design choices for the PC: coaxial and rectangular wave-guide couplers (Fig. 1).

The advantages of coaxial couplers are the higher compactness and the easy tuning of the match to the resonator and the beam by changing the penetration of the antenna into the resonator or beam pipe. In cryogenic systems the dynamic thermal RF losses of the inner conductor are cooled by 2/3 to room temperature or the 70 K intercept and not at the expensive 2 K or 4 K temperature. Multipacting can easily be suppressed by a high voltage bias on the isolated inner conductor (see also Section 6.3.). Mechanical arguments for the coaxial coupler design are the easy machining, assembly, and sealing of circular parts.

A disadvantage of coaxial couplers is the asymmetric field at the antenna which can cause a kick to the accelerated beam.

The advantages of rectangular wave-guide couplers are the lower surface electric field (1/4 compared to coaxial wave guides).

The disadvantages of rectangular wave-guide couplers are the difficulties in tuning the match to the resonator and the beam. In cryogenic systems one has to take into account the high thermal radiation through the big opening of the wave guide. The machining of big rectangular parts is more expensive, but in times of computer operated milling machines not as important as it used to be.

## 4 Coupler ports on the cavity

Coupler ports are located at cavity areas with high magnetic or electric field for efficient coupling. On normal-conducting (NC) cavities a convenient location is the cell equator for magnetic coupling. On the first SC cavities the coupler port was located at the equator like on normal-conducting cavities [7]. But the cavity performance was limited by multipacting in this area. All later SC cavity designs used a coupler port in the electric high field region on the beam tubes near to the cavity iris at the end cell of the resonator; see Fig. 2 [8, 9]. This leads to longer beam lines and accordingly to a smaller filling factor of the accelerator.

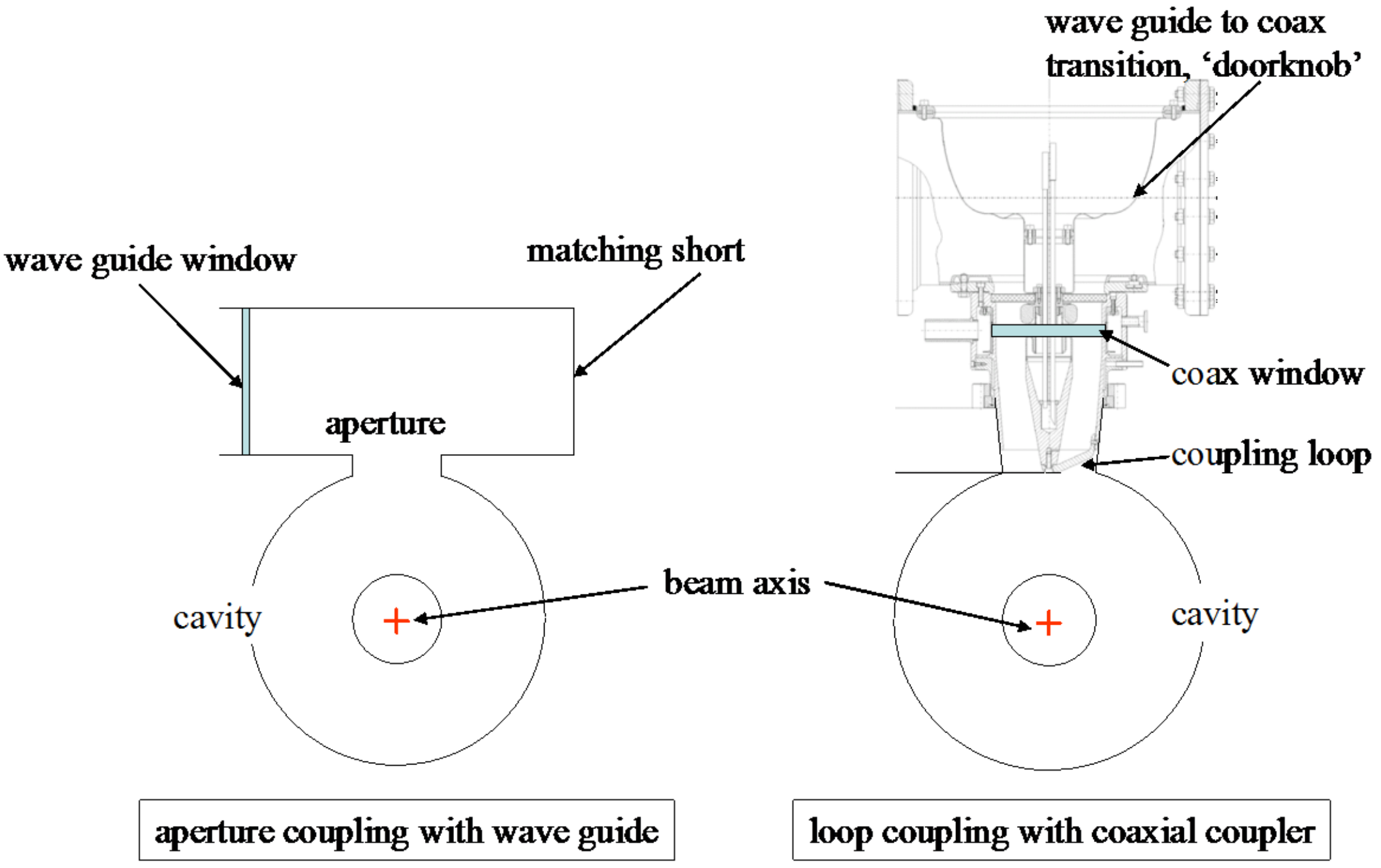

**Fig. 1:** Aperture coupling with a rectangular wave guide (left) and loop coupling with a coaxial coupler (right) on a normal-conducting cavity

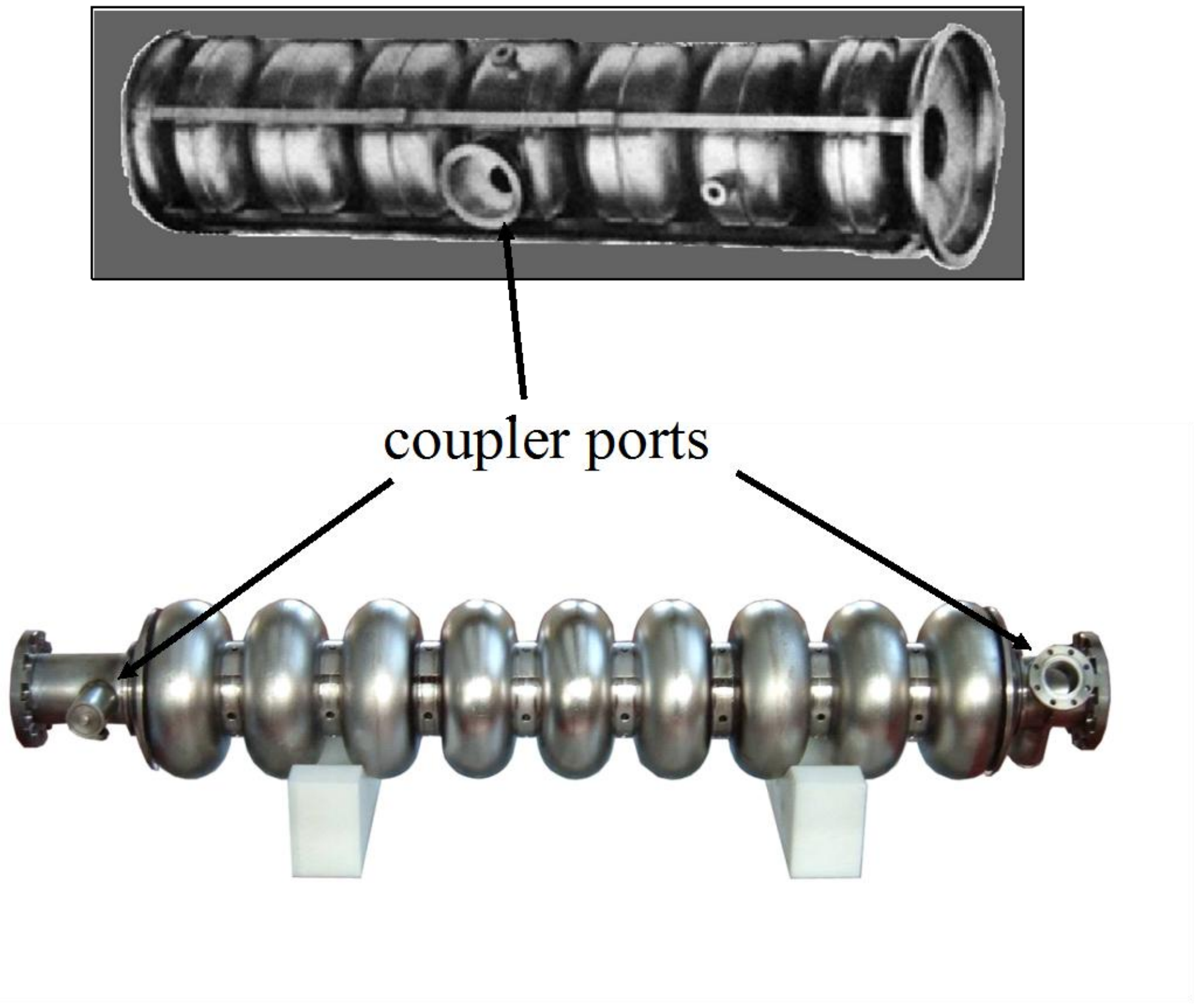

**Fig. 2:** Coupler ports on a superconducting cavity. Top: a subsection of the HEPL cavity (Stanford 1977) with the ports located on the equator (like NC cavities). Bottom: the TESLA cavity with the ports located on the beam line.

On the equator as well as at the beam tube a coupling to rectangular or to coaxial wave guide is possible and realized. Figures 3 and 4 show examples of PC for NC resonators and Figs. 5 [10] and 6 [11] for SC resonators.

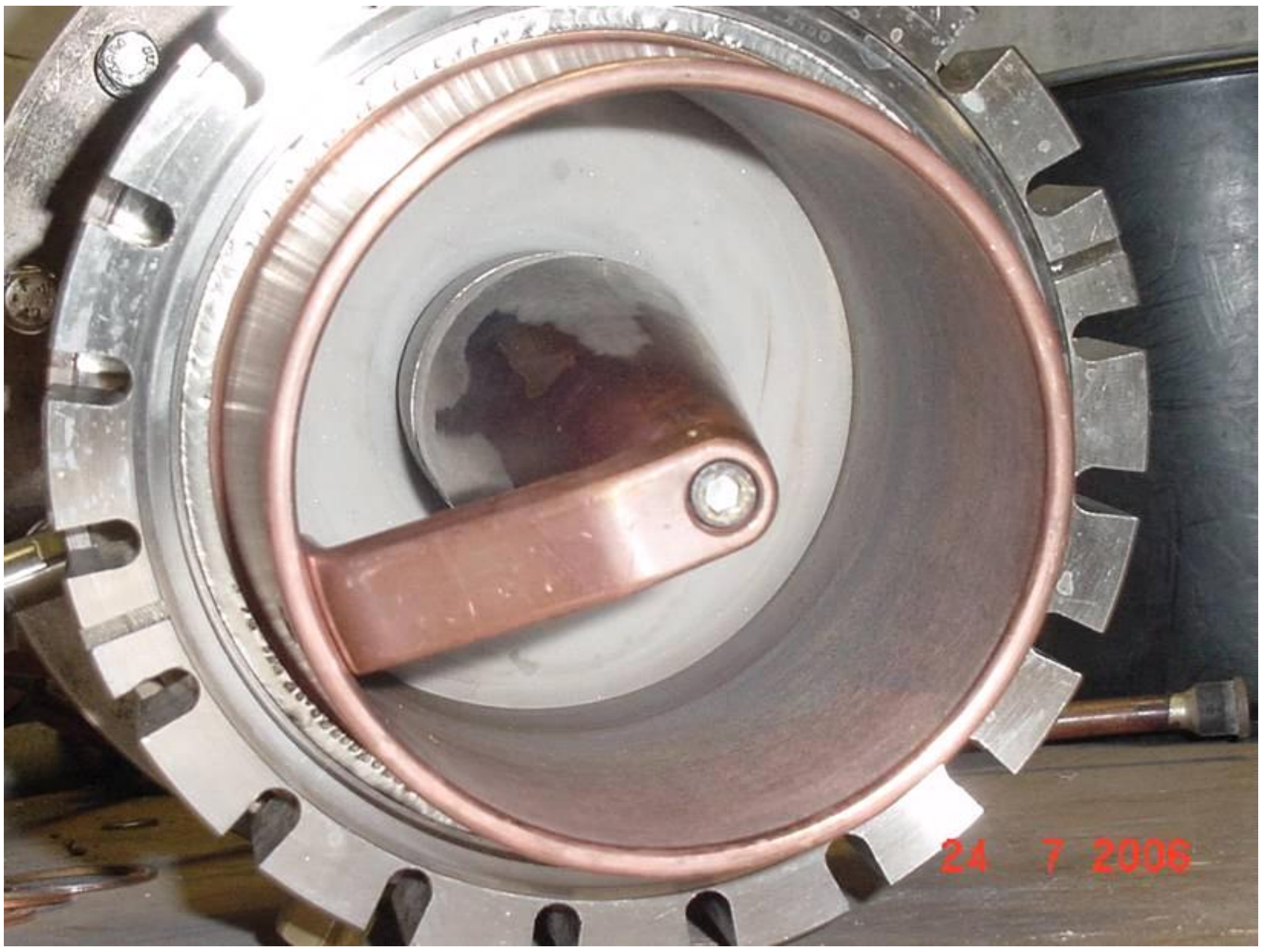

**Fig. 3:** Power coupler for the normal-conducting PETRA cavity at a frequency of 500 MHz and a maximum power of 250 kW, CW. The picture shows the coupling loop with the planar window ceramic in the coaxial line.

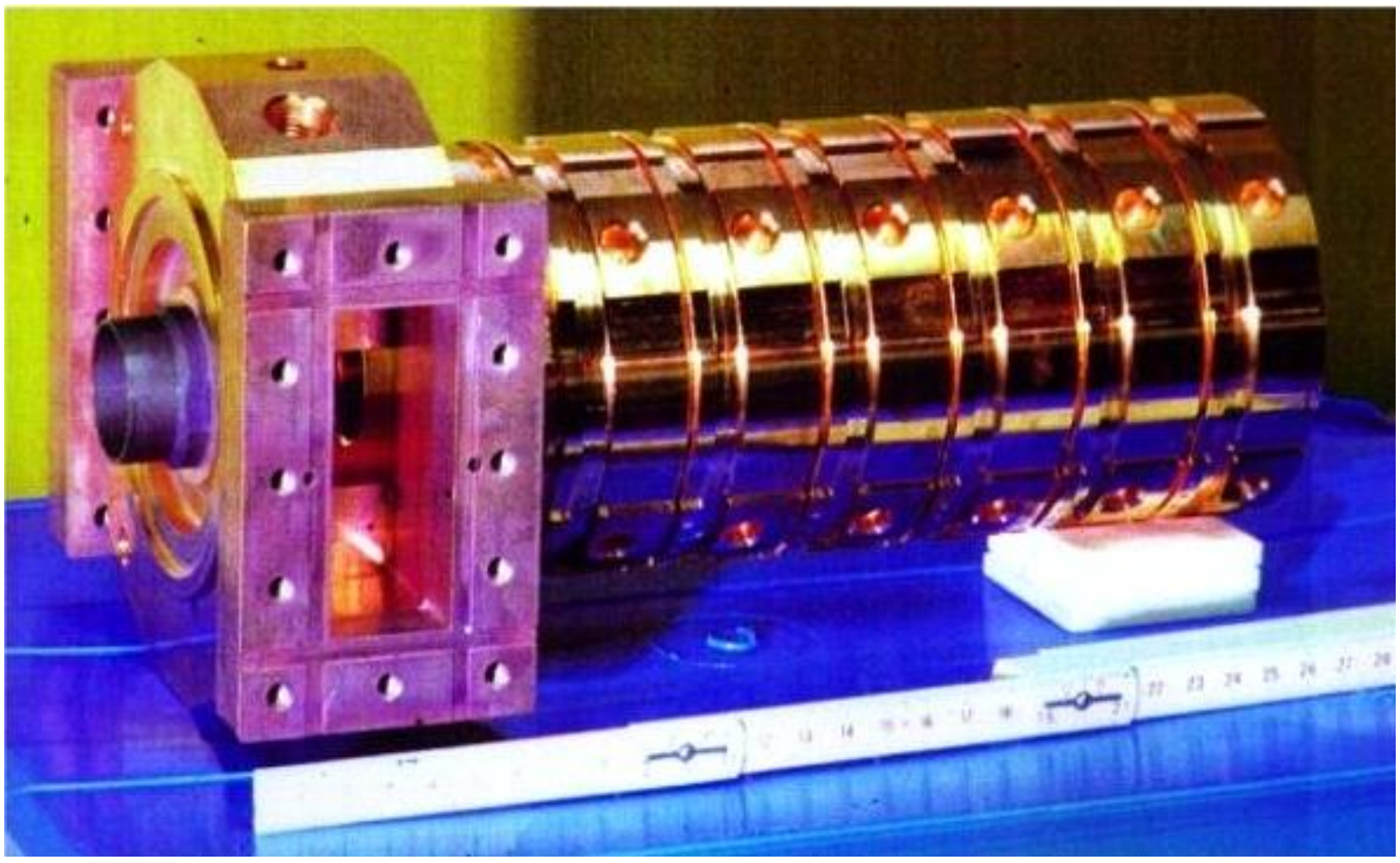

**Fig. 4:** The picture shows a normal-conducting S band structure with the wave-guide coupler ports on the left side. The frequency is 2.998 GHz and the maximum pulsed power is 100 MW.

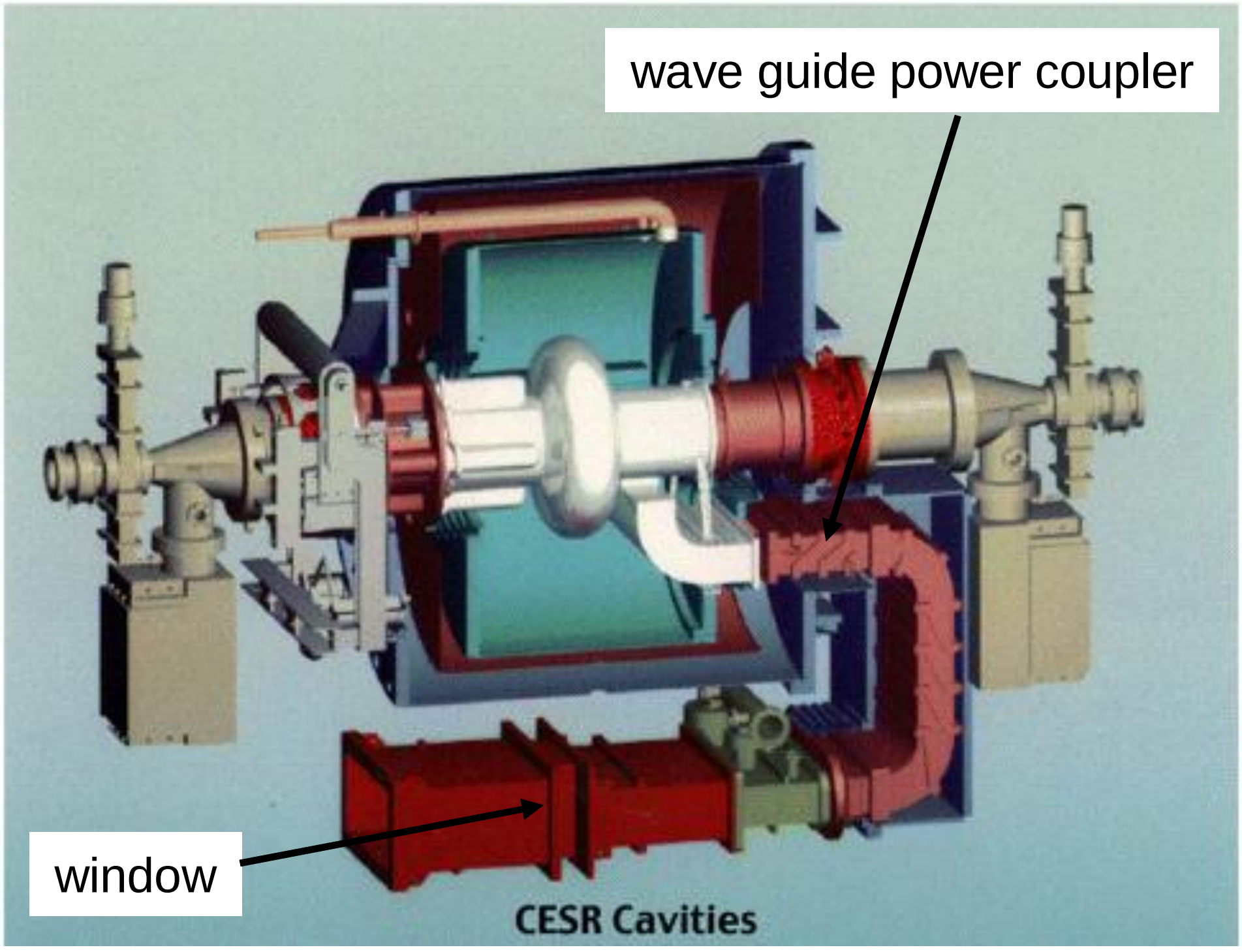

**Fig. 5:** The CESR-B superconducting cavity with the wave-guide coupling is shown. The ceramic wave-guide window is located at room temperature (RT).

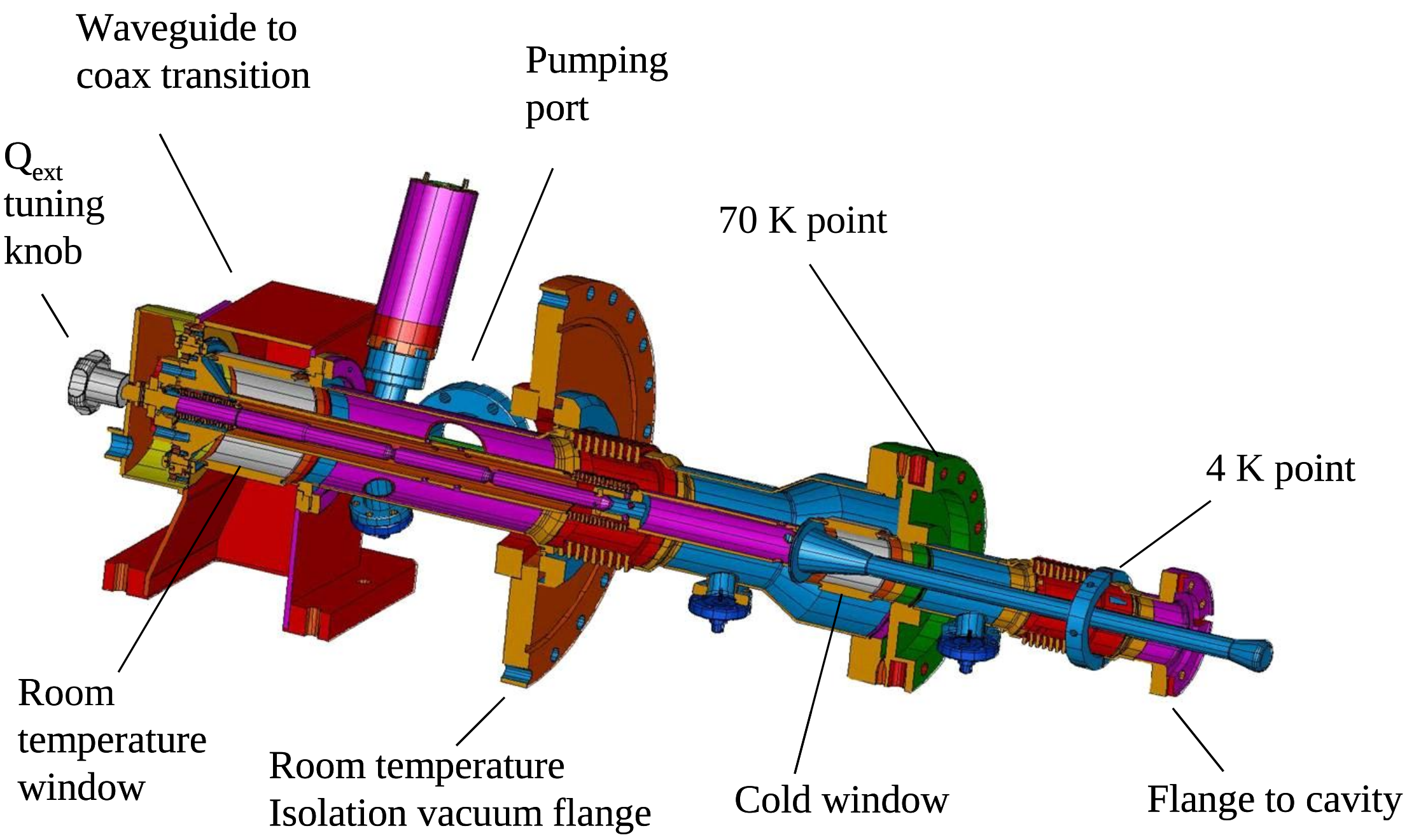

**Fig. 6:** Coaxial PC for the TESLA Test Facility with two cylindrical windows, one at room temperature and one at 70 K

For PC on SC resonators the integration into the cryogenic environment with the complicated thermal contractions during cool-down and warm-up has to be taken in account (Fig. 7). Therefore flexible parts like bellows have to be integrated into the coaxial or rectangular wave guides.

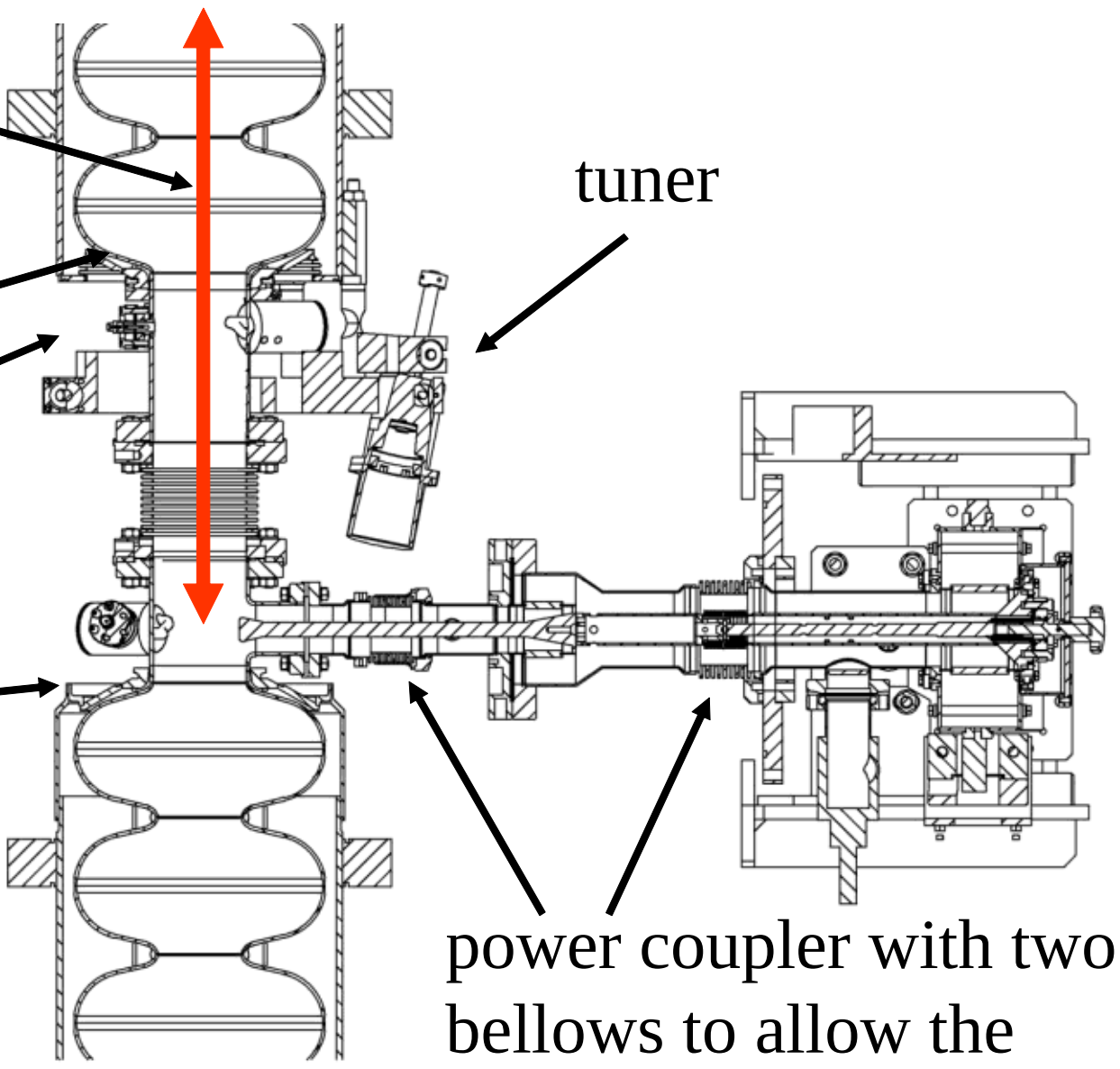

**Fig. 7:** The figure shows the different contractions and resulting movements of the cavity port during cool-down and warm-up on the example of the TESLA cryostat [12]

## 5 Two- vs. one-window coupler

For NC cavities the one-window solution is preferred. This allows a compact design and no extra vacuum system is needed. For SC cavities and low gradient applications (<15 MV/m) one-window couplers with the window at RT are also commonly used. For high gradient SC cavities (≥ 20 MV/m) the two-window design is common. Both, coupler and resonator have to be assembled in a very clean environment before other, unclean components can pollute the clean resonator, i.e., the window has to be close to the resonator for geometric restrictions at the cryostat. Since the window is at cryogenic temperature there is a need for a second window at RT with an intermediate vacuum to avoid gas condensation on the cold window. The second argument for a two-window design at SC cavities is the higher safety margin against window failure during operation.

## 6 RF design and simulations

The PC design specification has to respect the following parameters:

- frequency,
- maximum peak and average power,
- fixed or variable coupling,
- coaxial or rectangular wave guide,
- one or two windows,
- NC or SC resonator (cryostat design),
- window layout according to power handling, geometrical restrictions or available technology: disc, cylinder or cone.

For the RF design of the PC several finite-element simulation tools are available: SUPERFISH, URMEL, MAFIA™, HFSS™, CST MICROWAVE STUDIO® and others [13]. In the recent past a suite of 3D parallel finite-element based electromagnetic codes for accelerator modelling –- ACE3P ( Advanced Computational Electromagnetics 3P) have been developed at SLAC. It uses an unstructured grid for modelling geometries with large variation in dimensions and details with great realism [14]. Large systems like multi-resonator modules can be simulated.

## 6.1 Standing waves in power couplers at SC cavities

The RF losses in the SC resonator are negligible in comparison to the beam power. The input coupler is matched to the demand of beam power only. Therefore a sudden change or even loss of the beam current puts demanding conditions to the RF control circuit in keeping the resonator voltage constant and/or avoid large reflected RF power. For pulsed operation the coupler is operated at standing waves during the filling time and therefore has to carry an increased voltage compared to travelling waves. At the TTF3 PC the two windows are placed in the electric field minimum (Fig. 8).

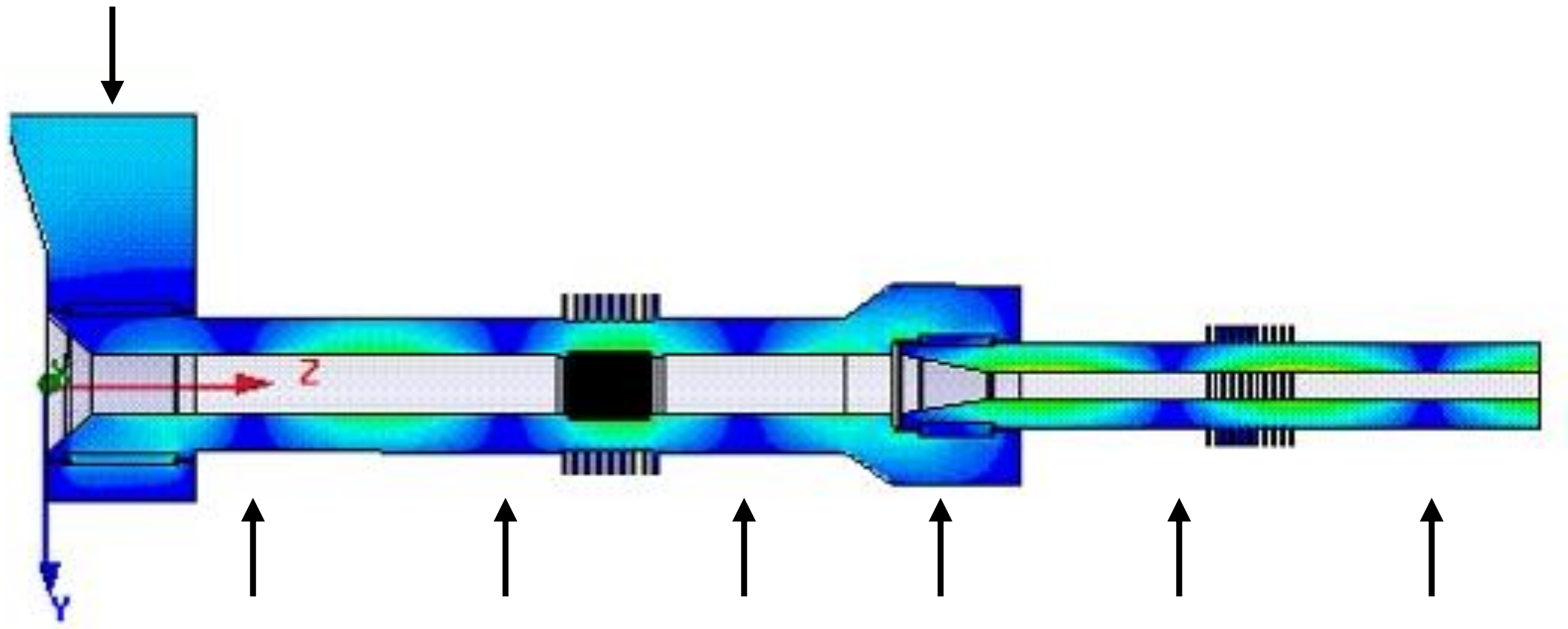

**Fig. 8:** RF simulation of the TTF3 PC during resonator filling time. The standing-wave electric field is shown. Both ceramic windows are placed in the electric field minimum.

## 6.2 Kick to the beam

When a coaxial coupler is located at the beam line, the asymmetric field at the antenna — beam pipe transition - causes an unwanted kick to the beam. At the Cornell ERL injector cryomodule this problem was solved by two symmetric couplers at opposite position. Also alternating coupler positions are possible. Another solution is a coaxial resonator in the beam line as used in the S-band linac at Darmstadt (Fig. 9).

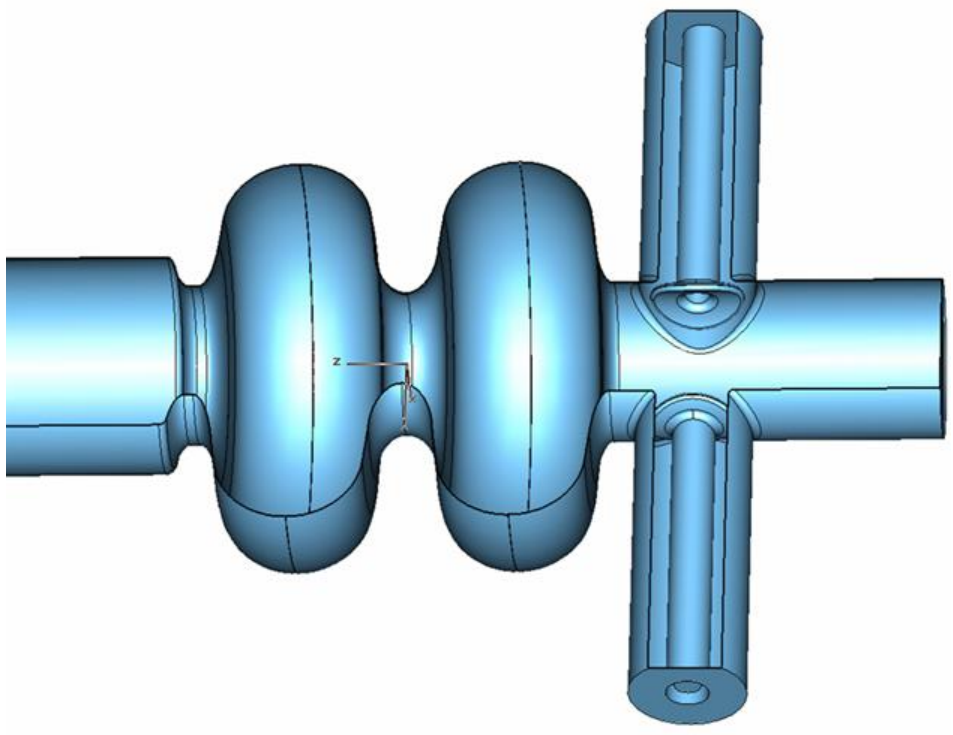

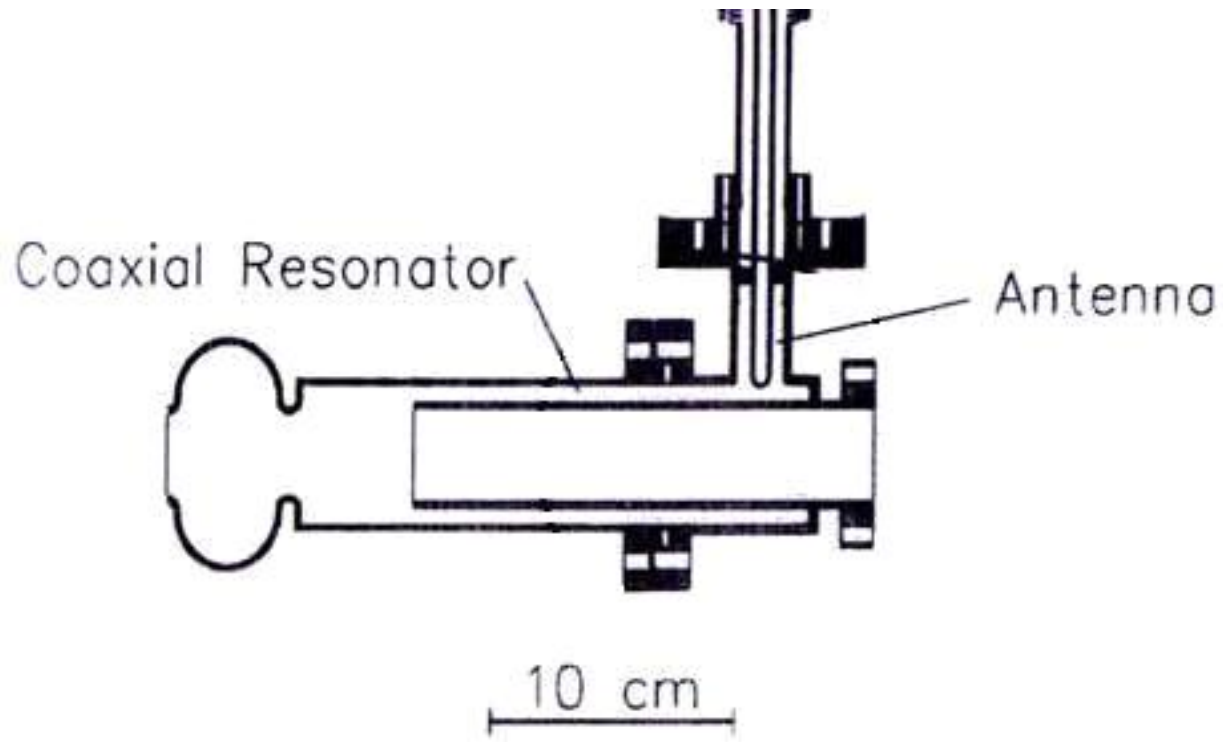

**Fig. 9**: Two PCs symmetrically placed at the beam line avoid an unwanted kick to the particle beam (left). A coaxial resonator in the beam line provides a symmetric coupling field (right).

### 6.3 Multipacting

During test or operation of a PC, electrons are accelerated by the electric field and desorb gas when impacting the RF surface. This can lead to conditions which increase the vacuum pressure and finally lead to an arc. Multipacting is a resonant multiplication of electrons caused by electron trajectories (1 point or 2 point) determined by RF field and geometry when the secondary electron emission coefficient (SEC) is higher than 1. In this way an avalanche of electrons can cause a breakdown of the electric field in the coupler. The order of the multipacting is defined by the number of the RF periods the electrons need for their travelling time. Lower orders are more stable and therefore it is more difficult to condition them away.

In a coaxial line the multipacting power level scales with frequency times diameter to the fourth power and linearly to the impedance (the ratio between the inner and outer coaxial diameter). The graph in Fig. 10 shows the analytically calculated multipacting thresholds for standing waves in coaxial lines for different power levels [15].

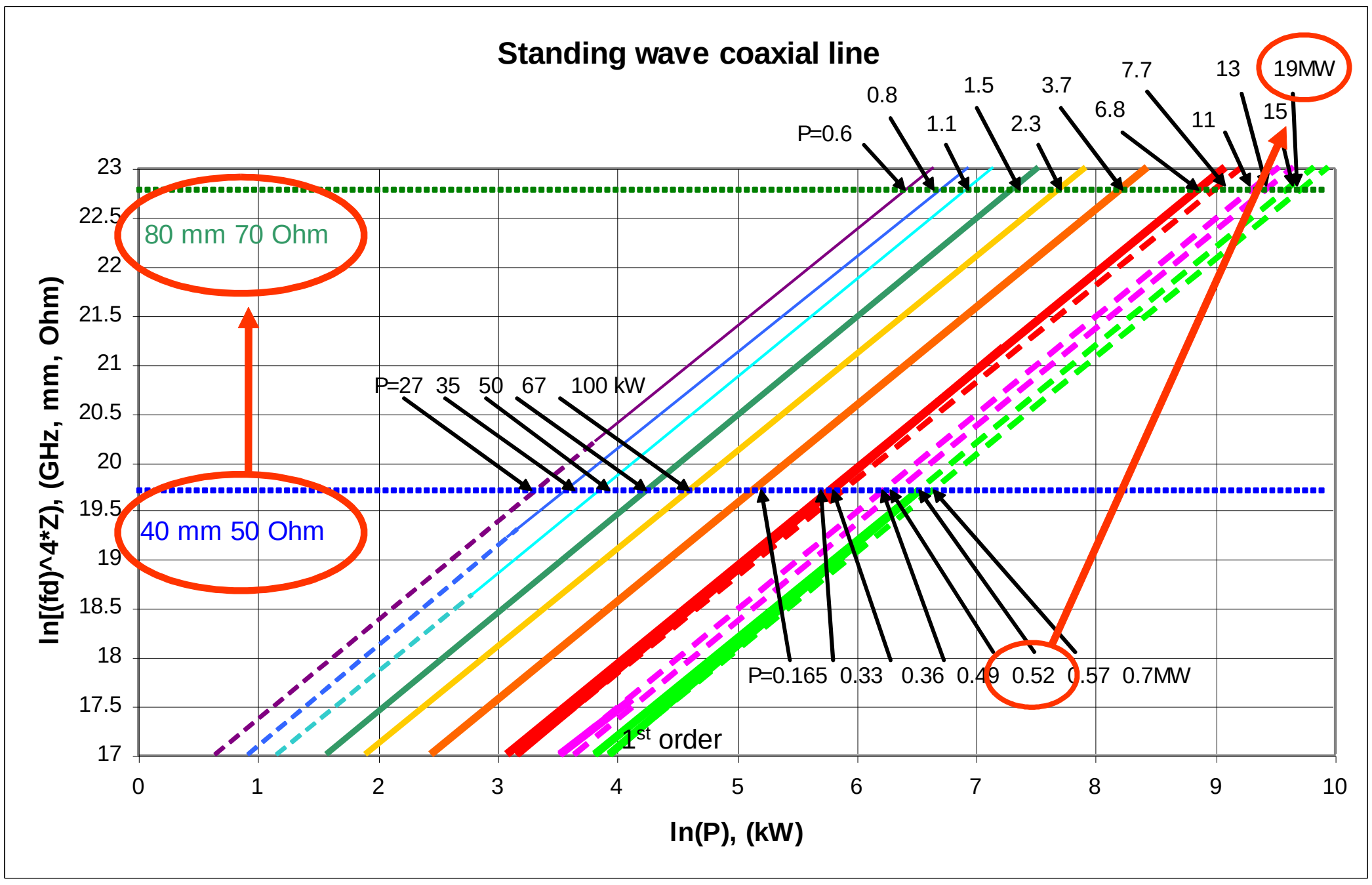

**Fig. 10:** Analytic calculations of multipacting thresholds in a coaxial line with standing waves. The y-coordinate shows the logarithmic dependency of the frequency, geometry and impedance, whereas on the x-coordinate the logarithmic power is drawn. The different coloured lines are: the first order multipacting is in green and the following second (purple), third (red), fourth (orange)…orders are on the left. As an example (in the bubbles) the increase from 40 mm to 80 mm coax diameter and the change from 50 to 70 ohm impedance are indicated. This geometric change shifts the most dangerous first-order multipacting threshold from 0.52 MW to 19 MW.

During the design phase the geometry has to be chosen to avoid multipacting. Increasing the diameter of a coaxial PC has a strong effect on shifting the multipacting levels to higher power levels beyond the operating power. Another precaution is the reduction of the SEC by coating the RF surfaces. Since $Al_2O_3$ has a high SEC of about eight at certain electron energies it is usually coated with TiN which has a SEC of about 1 at the same energies. A high SEC could also be caused by gas layers on the RF surfaces. A good cleaning and backing of the PC will help to reduce the multipacting. RF conditioning can clean the RF surfaces as well, but can be time consuming. A last remedy against multipacting is to destroy the resonant electrical conditions. This can be done by additional electrical or magnetic fields. At the coaxial coupler an electrical bias voltage on the inner conductor is an often used method. Figure 11 shows simulated power levels for multipacting depending on the bias voltage for the TTF3 PC [16].

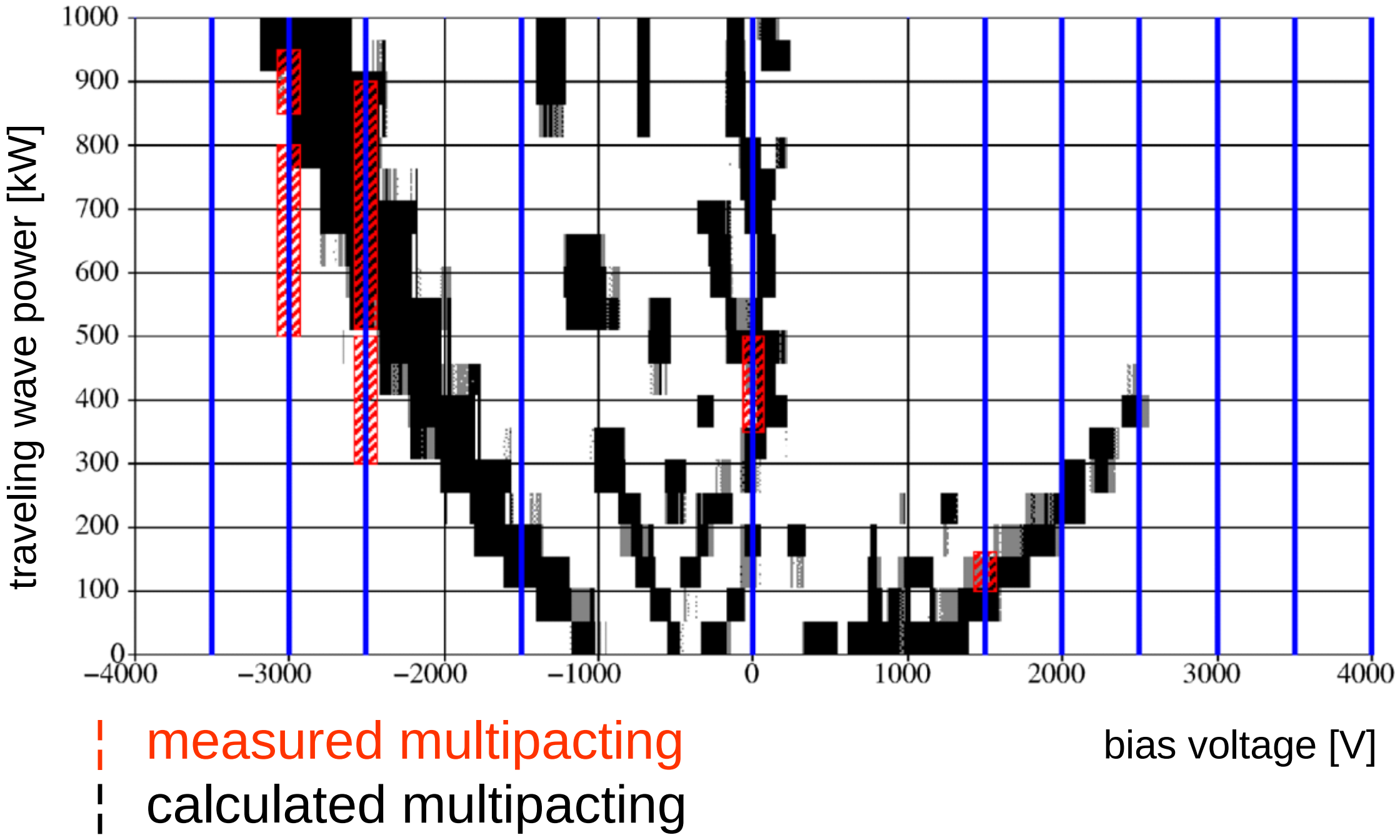

**Fig. 11:** Simulated multipacting thresholds in a travelling wave coaxial line are shown in dependency on the bias voltage on the inner conductor (black). In comparison the measurements are shown in red.

### 6.4 Thermal simulations

Thermal simulations can help to identify inhomogeneous heating of the ceramic and of other delicate parts of the PC. Figure 12 shows a thermal simulation of the 75 kW CW coaxial PC for the Cornell ERL injector cryomodule with the identified heating in the ceramic. A local air cooling was applied to reduce the thermal stress in the ceramic [17].

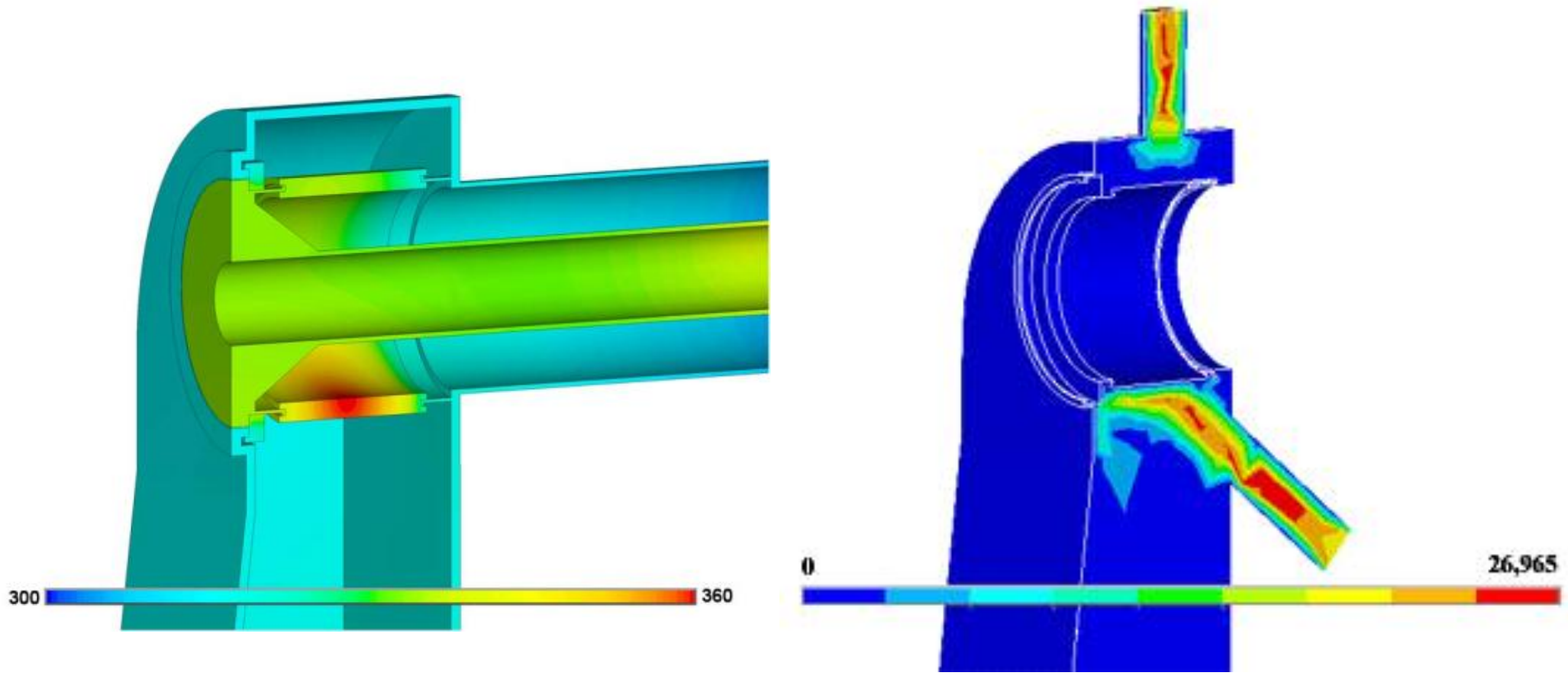

**Fig. 12:** Thermal simulations on the Cornell ERL PC war window. On the left side the thermal heating of the cylindrical ceramic is calculated. On the right side the cooling air flow simulation is shown.

For SC RF resonator systems the static and dynamic losses of the PC are important for the operating costs of the accelerator. A compromise between the favoured low thermal conductivity for low static cryogenic losses and the high electric conductivity for low dynamic losses has to be found. A good example of the optimization of the copper quality and thickness on the plated RF surfaces for different operating modes for the TTF3 PC is shown in Ref. [18].

## 7 Power coupler fabrication

A good RF design is a precondition for a reliably working PC. The fabrication involves many fabrication techniques and different materials. The number of machining, processing, and assembly steps is big and a failure during the fabrication can cause the loss of an expensive part. To realize a good coupler, the RF design has also to consider the fabrication, cleaning, handling, assembly and costs:

- use standard material qualities (316LN, Cu-OFHC, $Al_2O_3$),
- use standard sizes (tubes, bellows, flanges),
- use standard fabrication techniques,
- use fabrication techniques according to the abilities of the industries or workshops involved,
- decide on acceptable tolerances,
- clean handling during the fabrication is essential.

Close collaboration with the manufacturer as early as possible (already during the design phase) and during fabrication is a must.

### 7.1 Copper plating

PCs for SC resonators are very often made from stainless steel in order to reduce the static heat load for the cryogenic environment. For a high electrical conductivity the stainless steel has to be copper plated. Challenges for the copper plating process are the needed high electrical conductance for low RF losses and at the same time the small thickness for low thermal conductance and low static losses. In addition a good uniformity of thickness, especially on bellows, and low surface roughness with no blisters or stripping is required. The vacuum engineering asks for high cleanliness and low outgassing. Austenitic nickel chromium steel when exposed to air-oxygen atmosphere forms a very stable chromium oxide protecting the surface against corrosion. These oxide layers have a strong resistance against being coated. The activating processes for such materials must aim at removing the oxide layers and at forming a full covering, adhesion promoting intermediate layer [19].

Most of the copper layer properties can only be checked destructively [20]. Therefore the production and investigation of samples before and during fabrication is necessary.

### 7.2 Copper

PCs for NC resonators are normally fabricated from bulk high-purity copper because static losses are not an issue. The copper should not only have high electrical conductivity, but also must be 3D forged for high leak tightness on a small wall thickness.

### 7.3 Brazing

Many microwave tube industries prefer to braze fixtures and self-fixture assemblies. The advantage is that miscellaneous parts can be brazed at one time. The metallized ceramic must always be brazed to the joining parts. However, during braze under vacuum the ceramic has to be protected against evaporated metal. Brazing at high temperatures can cause copper grain size growth.

### 7.4 High austenitic steel

Electron and positron accelerators are very sensitive to perturbing magnetic fields. All components other than active steering elements near the particle beam have to be magnetically neutral. Therefore the steel used for PCs assembled to the resonator beam line has to be highly austenitic. The permeability is aimed to be $< 1.01$. The standards (e.g. EN 10088) allow a wide range of tolerances for the composition and the dosing tends to reduce the expensive materials. In case of doubt the chemical composition of the used stainless steel has to be verified [21].

### 7.5 Mechanical tolerances

Often the geometrical tolerances are set very tight in order to get the theoretical geometry of the RF design realized. Fabrication tolerances have a big impact on the fabrication methods and therefore on the price. Relaxed tolerances can make the fabrication much easier and also more economic. Detailed RF simulations can help to identify the most critical dimensions for the RF design of the PC and thus to find a good compromise.

### 7.6 Anti-multipacting coating

Because of the high SEC of the $Al_2O_3$ window it has to be coated to reduce the SEC. TiN is preferred because it has a low SEC and is a stable composition. Deposition processes are sputtering and evaporating of titanium. Ammonia or nitrogen is used to convert the Ti to TiN. Figure13 shows the DESY apparatus used for evaporating Ti on cylindrical windows [22].

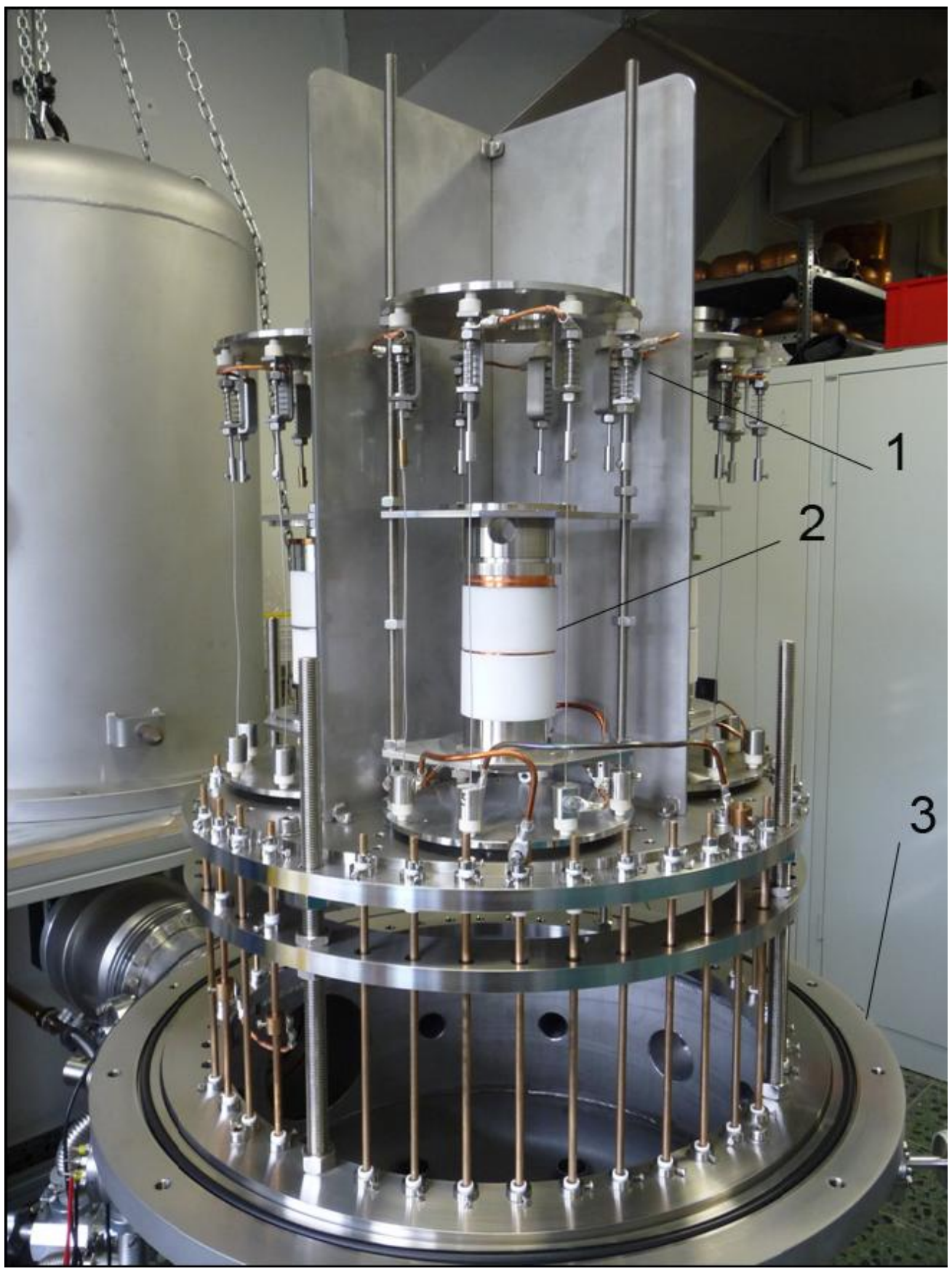

**Fig. 13:** Eight cylindrical window ceramics assembled to the TiN coating apparatus at DESY are shown. The flexible suspension of the Ti wire for Ti evaporation (1), the ceramics (2) and the vacuum vessel flange (3) can be seen.

## 8 RF test and conditioning

High-power coupler tests are needed as an acceptance test and for preconditioning prior to the operation on the resonator. Test stands are usually designed for two PCs and are operated at room temperature. A technical interlock is needed to protect the coupler and investigate the behaviour of the PC.

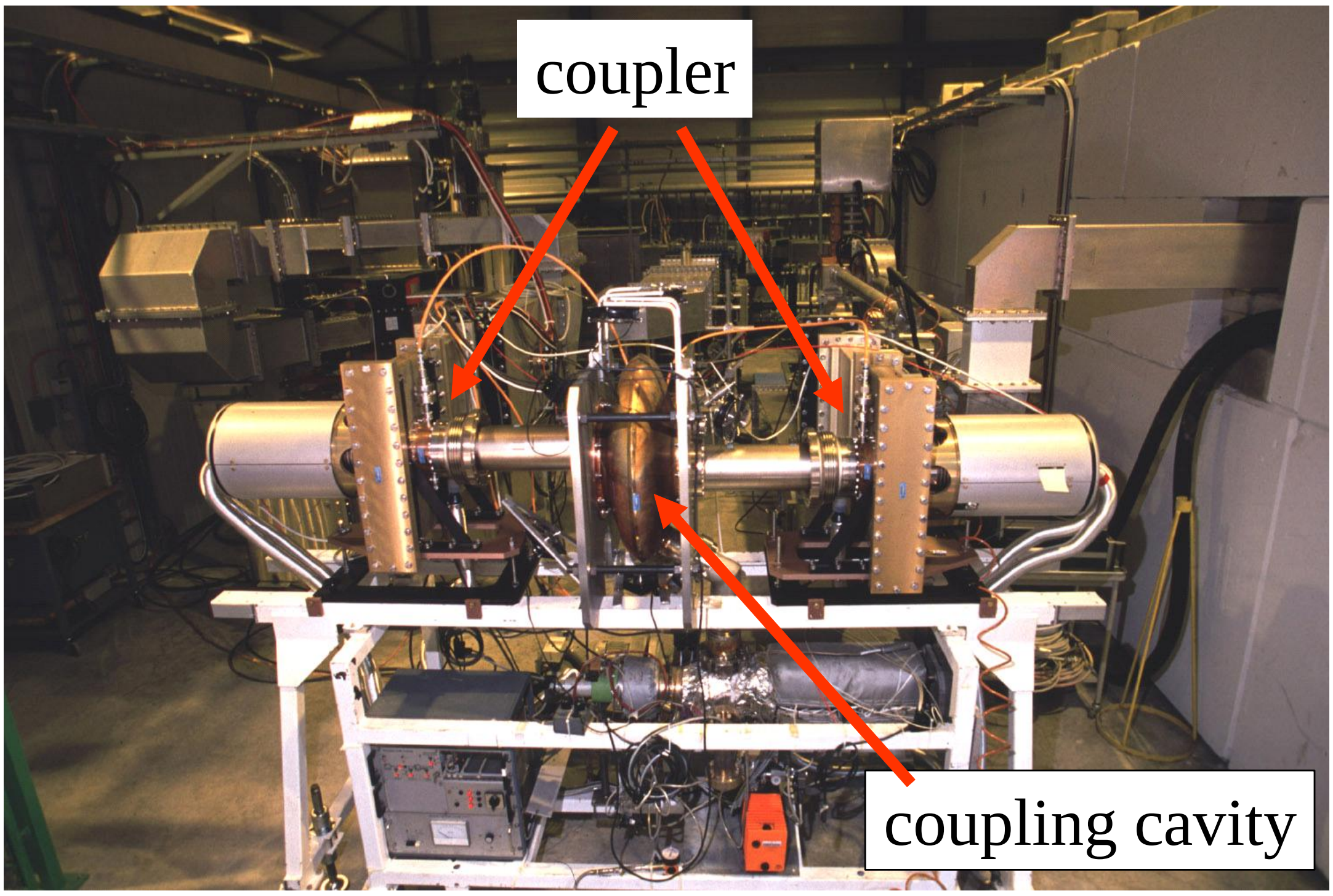

**Fig. 14:** The LHC power test stand. Two couplers, at the left and right side are connected to a coupling resonator in the middle.

RF conditioning, also called ageing of a PC, is the controlled desorption of absorbed gases by accelerated ions and electrons from the RF surfaces. A compromise must be found between conditioning speed and sparking risk. Travelling waves clean the entire RF surfaces, with standing waves only the high field part is conditioned. During PC operation on SC cavities the cold surfaces collect gas after a certain period of operation. This might necessitate a reconditioning.

The testing and conditioning procedure consist of: starting at low power, short pulses and low repetition rate and proceeding to high power, long pulses and high repetition rate. To avoid RF breakdown the power rise is limited by thresholds of coupler vacuum, measurements of charged particles like e- and light in the coupler. Fast vacuum feedback loops, called ‘analogue processing’ are used to keep the power level close to the vacuum thresholds and guarantee an efficient and fast processing [23]. At KEK the bias voltage on inner coax was used to process the multipacting levels. Controlled discharge processing with argon or helium is another possible processing method.

The technical interlock system protects the coupler against discharge and other degradations. There is usually a hardware interlock consisting of the following channels: vacuum read out, e- pick up, light detectors in vacuum and on the air side, temperature measurement on windows, and reflected power measurements. A software interlock including all the above mentioned channels is used with slightly tighter thresholds than the hardware interlock in order to avoid long RF off time. Figure 14 shows a typical RF test stand.

## 9 Handling before and after conditioning

In order to avoid oxidation and contamination of the PC parts the storage is always done under dry nitrogen. PCs for SC cavities are cleaned according to the ISO 4 standard with ultra pure water and are assembled in a clean room. After the test stand assembly a vacuum backing is essential for fast conditioning.

The goal is to preserve the test stand conditioning effect to avoid longer additional conditioning in the accelerator. Therefore the PC parts disassembly from test stand and assembly to the resonator and cryogenic module are done under clean conditions. During storage of the PC the RF surfaces should always be protected against contamination.

## Acknowledgement

I wish to thank all colleagues working in the field of RF high-power couplers.